\documentclass[12pt]{article}

\usepackage{epsfig}
\usepackage{amsmath}
\usepackage{amsfonts}
\usepackage{amssymb}

\title{{\bf Chiral Symmetry Breaking in a Soft-Wall Model of AdS/QCD}}

\author{{JOSEPH I. KAPUSTA$^*$ and THOMAS M. KELLEY}\\
{\it School of Physics and Astronomy, University of Minnesota}\\
{\it Minneapolis, MN 55455, USA}\\
{\it $^*$E-mail: kapusta@physics.umn.edu}\\
{\it kelley@physics.umn.edu \vspace*{0.1in}}\\
TONY GHERGHETTA\\
{\it School of Physics, University of Melbourne}\\
{\it Victoria 3010, Australia}\\
{\it E-mail: tgher@unimelb.edu.au}}

\begin{document}

\maketitle

\begin{abstract}
\noindent We incorporate chiral symmetry breaking in a soft-wall version of 
the AdS/QCD model by using a modified dilaton profile and a quartic term in the bulk scalar potential.  This allows one to separate the dependence on spontaneous and explicit chiral symmetry breaking.  The resulting mass spectra in the scalar, vector and axial-vector sectors compares favorably with the respective QCD resonances. 
\end{abstract}


\section{Introduction} 
\label{secIntro}

The AdS/CFT correspondence \cite{Maldacena:1997re, Gubser:1998bc, Witten:1998qj, Klebanov:1999tb} provides a remarkable mathematical tool with which to understand strongly-coupled gauge theories. Calculations performed on the gravity side can be reinterpreted as due to nonperturbative effects on the field theory side. The theory of the strong interactions, quantum chromodynamics (QCD), is a strongly-coupled gauge theory at low energies, and therefore a natural candidate with which to apply the gauge/gravity correspondence.  This had led to a bottom-up approach, commonly known as AdS/QCD \cite{Erlich:2005qh, DaRold:2005zs}, which relates QCD to a five-dimensional (5D) gravity theory. This simple model captures the essential features of the low-lying meson spectrum.

To incorporate more realistic features of the excited states, the AdS/QCD model can be modified to include a dilaton with a quadratic profile \cite{Karch:2006pv}. While the linear radial spectrum is indeed produced in this soft-wall version of the AdS/QCD model, the form of chiral symmetry breaking is not QCD-like. In particular, the bulk scalar field, dual to the quark bilinear operator ${\bar q} q$, whose vacuum expectation value is responsible for spontaneous chiral symmetry breaking, does not allow the spontaneous and explicit breaking to be independent. Moreover, chiral symmetry is restored for the highly excited states, a phenomenological feature that is not supported in the QCD mass spectrum \cite{Shifman:2007xn}. 

We \cite{us} modify the existing soft-wall version of the AdS/QCD model in order to incorporate these two phenomenological features of QCD. This is done by adding a quartic term to the bulk scalar potential and changing the dilaton profile. By assuming that the bulk scalar field contains the desired limiting behavior for non-restoration of chiral symmetry \cite{Klebanov:1999tb, Shifman:2007xn}, we derive a new dilaton background profile. The extra parameter introduced by the quartic term decouples the quark mass from the chiral condensate, thereby allowing for spontaneous and explicit chiral symmetry breaking to occur independently.

\section{The Model} 

We consider a modified version of the soft-wall AdS/QCD model first introduced in Ref. \cite{Karch:2006pv} and investigated further in Refs. \cite{Evans:2006ea,Grigoryan:2007my,Kwee:2007nq,Cherman:2008eh,Colangelo:2008us,Batell:2008zm,Huang:2007fv}. 
The background geometry is assumed to be 5D AdS space with the metric
\begin{equation}
ds^{2}=g_{MN}dx^{M}dx^{N}=a^2(z)\left( \eta_{\mu\nu}dx^{\mu}dx^{\nu}+dz^{2}\right)~,
\end{equation}
where $a(z)=L/z$ is the warp factor, $L$ is the AdS curvature radius and the Minkowski metric $\eta_{\mu\nu}={\rm diag}(-1,+1,+1,+1)$. The conformal coordinate $z$ has a range $0\leq z < \infty$. To obtain linear trajectories, Ref. \cite{Karch:2006pv} also introduced a background dilaton field
$\phi(z) = \lambda z^2$ where $\lambda$ sets the mass scale for the meson spectrum. 

To describe chiral symmetry breaking in the meson sector the 5D action contains
SU(2)$_L\times$ SU(2)$_R$ gauge fields and a bifundamental scalar field $X$. As suggested by Ref. \cite{Karch:2006pv}, we add a quartic term in the potential $V(X)$ of our 5D action,
\begin{equation} \label{action1}
S_{5}=-\int d^{5}x \sqrt{-g}\,e^{-\phi(z)}{\rm Tr}\left[|D X|^{2}+ m_{X}^{2} |X|^{2}-\kappa |X|^{4}+\frac{1}{4 g_{5}^{2}}(F_{L}^{2}+F_{R}^{2})\right],
\end{equation}
where $m_{X}^{2} = - 3/L^{2}$, $\kappa$ is a constant, $g_5^2=12\pi^2/N_c$, 
and the covariant derivative is $D^M X=\partial^M X-i A_L^MX+iXA_R^M$. To describe the vector and axial-vector mesons we simply transform to the vector (V) and axial-vector (A) fields where $V^{M}=\frac{1}{2}(A_{L}^{M}+A_{R}^{M})$ and $A^{M}=\frac{1}{2}(A_{L}^{M}-A_{R}^{M})$.

The scalar field $X$, which is dual to the operator $\bar{q} q$, is assumed to obtain a $z$-dependent vacuum expectation value (VEV), 
\begin{equation}
\label{xvev}
\langle X \rangle \equiv \frac{v(z)}{2}
\left( \begin{array}{cc} 
  1 & 0 \\
  0 & 1  
\end{array} \right),
\end{equation}
which breaks the chiral symmetry SU(2)$_L\times$SU(2)$_R to $SU(2)$_V$. Assuming (\ref{xvev}) we obtain a nonlinear equation for $v(z)$,
\begin{equation}
\partial_z(a^3 e^{-\phi} \partial_z v(z))-  a^5 e^{-\phi} (m_X^2 v(z)-\frac{\kappa}{2} v^3(z))=0.
\label{VEVequation}
\end{equation}
When $\kappa=0$, the solution of (\ref{VEVequation}) which leads to a finite action in the limit $z\rightarrow\infty$ is given 
by~\cite{Karch:2006pv, Colangelo:2008us}
\begin{equation} \label{equk0}
v(z) = m_q \frac{z}{L} \, 
U\left( \frac{1}{2},0,\lambda \frac{z^2}{L^2} \right),
\end{equation}
where $U(a,b,y)$ is the Tricomi confluent hypergeometric function.

As expected from the AdS/CFT dictionary established in Refs. \cite{Klebanov:1999tb,Witten:1998qj}, as $z\rightarrow 0$ the VEV should take the form
\begin{equation}
v(z)= \frac{z}{L} \left( m_q \zeta + \frac{\Sigma}{\zeta} z^2 + 
\cdot\cdot\cdot \right) \, .
\label{smallV}
\end{equation}
Here $m_q$ is the quark mass, $\Sigma \equiv \langle \bar{q} q \rangle$ is the chiral condensate, and $\zeta$ is the normalization parameter introduced in Ref. \cite{Cherman:2008eh}. For fixed values of $m_{q}$ and $\Sigma$, the introduction of $\zeta$ still satisfies the Gell-Mann-Oakes-Renner relation, $m_\pi^2 f_\pi^2=2 m_q\Sigma$.  In actuality the normalization $\zeta$ is not a free parameter but is determined by QCD, as shown in Ref. \cite{Cherman:2008eh}, to be $\zeta=\sqrt{3}/g_5$.  Expanding the solution (\ref{equk0}) in the small $z$ limit leads to $\alpha\propto m_q$ and $\beta\propto \Sigma \propto m_q$. Thus in the limit $m_q \rightarrow 0$ the model eliminates explicit and spontaneous chiral symmetry breaking in contradiction with QCD. It will be seen that the introduction of a quartic term in the potential $V(X)$ avoids the dependence of the chiral condensate on the quark mass encountered in Refs. \cite{Karch:2006pv,Kwee:2007nq,Colangelo:2008us}. 

The solution (\ref{equk0}) has an asymptotic limit $v(z)\rightarrow$ constant for large values of $z$.  This asymptotic behavior suggests chiral symmetry restoration in the mass spectrum, a phenomenon not supported in QCD (although speculation continues on whether such a restoration indeed exists \cite{Cohen:2005am,Wagenbrunn:2006cs}). As noted in Ref. \cite{Shifman:2007xn} the highly excited mesons exhibit seemingly parallel slopes signifying that chiral symmetry is not restored. In order to incorporate this behavior $v(z)$ must behave linearly as $z$ becomes large, $v(z\rightarrow\infty) \sim z$ causing the mass-squared difference between vector and axial-vector resonances to approach a constant as $z\rightarrow\infty$. By including a quartic term and requiring $v(z)$ to have this linear asymptotic behavior we aim to incorporate these QCD-like characteristics into the soft-wall model.

Instead of solving for $v(z)$ directly, we assume the VEV asymptotically behaves as expected, namely Eq. (\ref{smallV}) and
\begin{equation}
v(z\rightarrow\infty) = \frac{\gamma}{L} z \, , 
\label{largeV}
\end{equation}
and then solve for the dilaton $\phi(z)$ using (\ref{VEVequation}), which becomes
\begin{equation}
\label{phieqn}
\phi'(z)=\frac{1}{a^3 v'(z)}\left[\partial_z(a^3 v'(z))-a^5 (m_X^2 v(z)- \frac{\kappa}{2} v^3(z))\right],
\end{equation}
where the prime denotes the derivative with respect to $z$. Given the required behavior (\ref{smallV}) and (\ref{largeV}) we can uniquely determine the dilaton profile up to a constant. With this procedure the two sources of chiral symmetry breaking decouple while simultaneously allowing for linear trajectories in the meson spectrum.

A particularly simple parametrized form for $v(z)$ that satisfies (\ref{smallV}) and (\ref{largeV}) is
\begin{equation}
v(z) = \frac{z}{L}(A + B \tanh{C z^2}), \label{arcv}
\end{equation} 
The parameters $\gamma$, $A$, $B$ and $C$ can be expressed in terms of the input parameters $m_q,\Sigma,\lambda,\kappa$ as 
\begin{eqnarray}
\gamma &=& \sqrt{\frac{4 \lambda}{\kappa}}, \label{padeC}\\
A &=& \frac{\sqrt{3}m_q}{g_5}, \label{arcA} \\
B &=& \gamma - A, \label{arcB} \\
C &=& \frac{g_5 \Sigma}{\sqrt{3}B}. \label{arcC}
\end{eqnarray}
The parameter $\lambda$ is determined by the average slope of the radial trajectories of the scalar, vector, and axial-vector mesons for radial quantum numbers $n \ge 3$. Its value was determined to be $\lambda = 0.1831$ GeV$^2$, as explained in the next section.  The quark mass, quark condensate, pion decay constant, and pion mass are all related by the Gell-Mann-Oakes-Renner relation, $f_{\pi}^2 m_{\pi}^2 = 2 m_q \Sigma$.  This relation holds in this model as a natural consequence of chiral symmetry \cite{Erlich:2005qh}.  We use the measured values of $f_{\pi}=92.4$ MeV and $m_{\pi}=139.6$ MeV, and adjust the quark mass to reproduce the input value of $f_{\pi}$ from a solution to the axial-vector field equation in Section~\ref{secPion} for a given value of $\kappa$.  The parameter $\kappa$ essentially controls the mass splitting between the vector and axial-vector mesons.  It is determined to be $\kappa = 15$ by a best fit to the radial spectra of the axial-vector mesons.  This results in $m_q = 9.73$ MeV and therefore $\Sigma =(204.5$ MeV)$^3$.  The inferred value of the quark mass is consistent with an average of the up and down quark masses as summarized in the Review of Particle Physics \cite{pdg} as appropriate at the hadronic energy scale.
  
\begin{figure}
\begin{center}
\includegraphics[scale=0.40,angle=90]{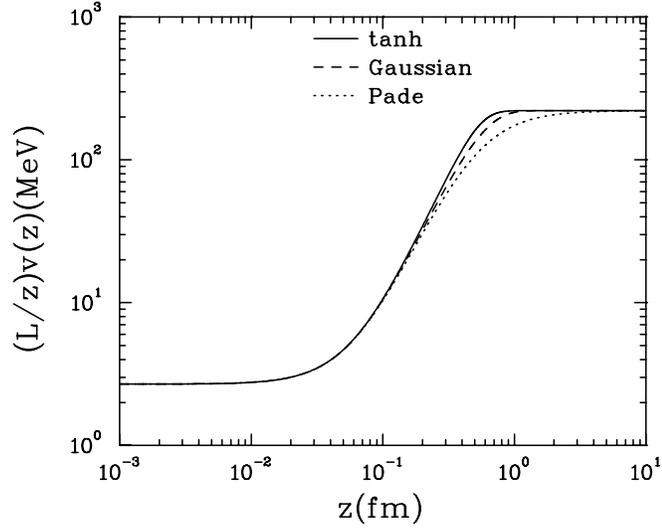}
\caption{\textsl{A plot of $v(z)/z$ for various parameterizations fitted to the mass spectra. The best fit to the mass spectra is obtained with the tanh form (\ref{arcv}).}}
\label{vevfig}
\end{center}
\end{figure}

\begin{figure}
\begin{center}
\includegraphics[scale=0.40,angle=90]{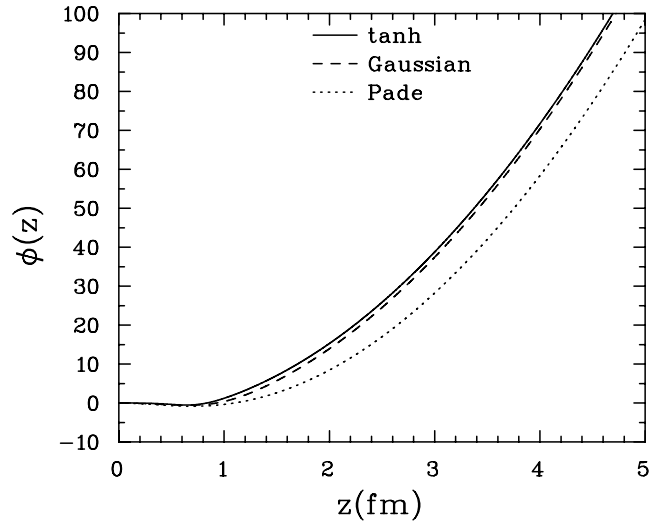}
\caption{\textsl{The dilaton profile $\phi(z)$ resulting from the various parameterizations of $v(z)$. The best fit to the
meson spectra occurs with the tanh parameterization (\ref{arcv}). For $z\lesssim1$ the behavior deviates from the quadratic asymptotic form.}}
\label{dilatonfig}
\end{center}
\end{figure}

Other parameterizations of $v(z)$ which lead to qualitatively similar behavior as that required in (\ref{smallV}) and (\ref{largeV}) can also be used. Other forms for $v(z)/z$ include $(a_1 + a_2 z^2)/(1 + a_3 z^2)$ [Pade] and $c_1 - c_2 \exp(-c_3 z^2)$ [Gaussian]. These forms were all studied but the best results were found to be obtained using the form (\ref{arcv}).  These parameterizations are shown in Fig. \ref{vevfig} using the above parameters.  The resulting dilaton profile $\phi(z)$ is shown in Fig. \ref{dilatonfig}.

\section{Meson Mass Spectra}

\begin{figure}
\begin{center}
\includegraphics[scale=0.40,angle=90]{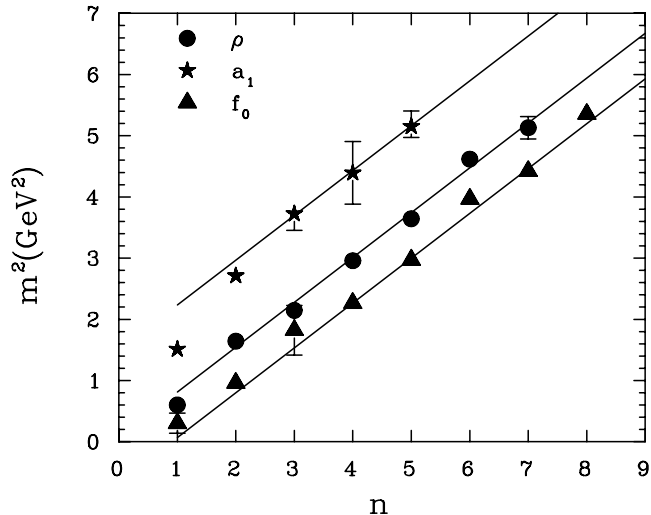}
\caption{\textsl{A straight-line fit to the measured scalar, vector and axial-vector mass spectra for $n \ge 3$ used to determine the dilaton mass parameter
$\lambda$. }}
\label{combinedplot}
\end{center}
\end{figure}

The soft wall model can be used to fit the observed scalar, vector, and 
axial-vector meson mass spectra \cite{pdg,Bugg,Bertin,bl}.  All but one of the included states are listed in the Review of Particle Physics (RPP) \cite{pdg}.  For an explanation of the masses used see Ref. \cite{us}.  A straight line is fitted to the $m^2$ versus $n$ plot with $n \ge 3$ for all three mesons, assuming the same slope $4\lambda$ but different intercepts, namely $m^2_n = 4\lambda n + m^2_0$.  The results are: $\lambda = 0.1831 \pm 0.0059$ GeV$^2$, $m_{V,0}^2 = 0.0806 \pm 0.0104$ GeV$^2$, $m_{A,0}^2 = 1.5023 \pm 0.0366$ GeV$^2$, and $m_{S,0}^2 = -0.6634 \pm 0.0038$ GeV$^2$.  See Fig. \ref{combinedplot}.  We use this value of $\lambda$ in our model calculations.

\subsection{Scalars} 
\label{secScalar}

\begin{figure}
\begin{center}
\includegraphics[scale=0.40,angle=90]{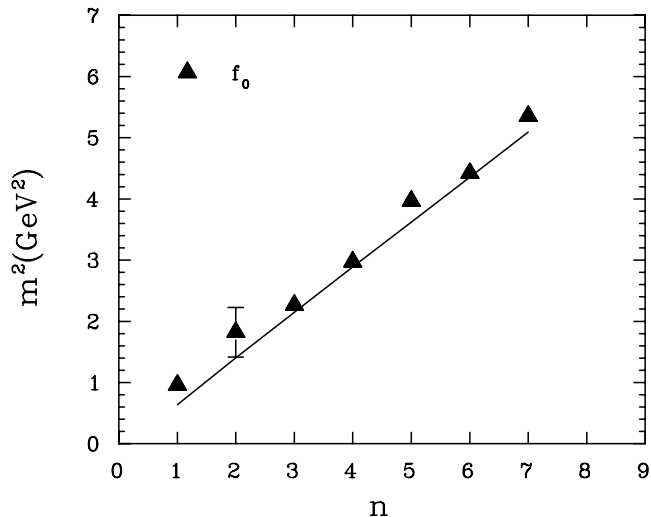}
\caption{\textsl{Comparison of the predicted scalar mass eigenvalues using the tanh form (\ref{arcv}) of $v(z)$ (solid) with the QCD $f_{0}$ scalar mass spectrum with the $\sigma$ meson removed \cite{pdg}.}}
\label{ScalarMassesNoSigma}
\end{center}
\end{figure}

Introducing a quartic term in the Lagrangian causes the scalar excitations to couple with their own VEV, giving a modified equation of motion unlike those in Refs. \cite{DaRold:2005zs,Colangelo:2008us}. The equation of motion can be brought into a Schr\"{o}dinger-like form
\begin{equation} \label{schroS}
-\partial_z^{2}s_n(z)+\left(\frac{1}{4}\omega_s'^2-\frac{1}{2}\omega_s''-\frac{3}{2}\frac{L^2}{z^2} \kappa v^2(z)-\frac{3}{z^2}\right)s_n(z)=m_{S_n}^{2}s_n(z).
\end{equation}
where $\omega_s = \phi(z) + 3\log{z}$.  To solve this equation we implement a shooting method in which (\ref{schroS}) is solved for various values of $m_{S_n}$. The eigenvalues are then those mass values that produce a solution for $S_{n}(z)$ that is bounded as $z\rightarrow\infty$.  Applying the shooting method to (\ref{schroS}) with the boundary conditions $\lim_{z_0\rightarrow 0} s_n(z_0)=0$, $ \partial_z s_{n}(z\rightarrow\infty)=0$ produces a scalar mass spectrum which has the correct slope but too large magnitude. This could well be a failure of this specific model.  However, it has been argued\cite{LinSerot} that the $\sigma$ meson is not truly a particle but only an enhancement in the $\pi$-$\pi$ cross section, and that the phase shift does not reach 90$^o$.  Removing the $f_0(550)$ would shift all the higher masses to the left by one unit of $n$, resulting in a better fit to the model.  See Fig. \ref{ScalarMassesNoSigma}.  An obvious extension of this work would be to include strange quarks and glueballs and to determine the mixing among the resulting scalar states.

\subsection{Vectors} 
\label{secVector}

\begin{figure}
\begin{center}
\includegraphics[scale=0.40,angle=90]{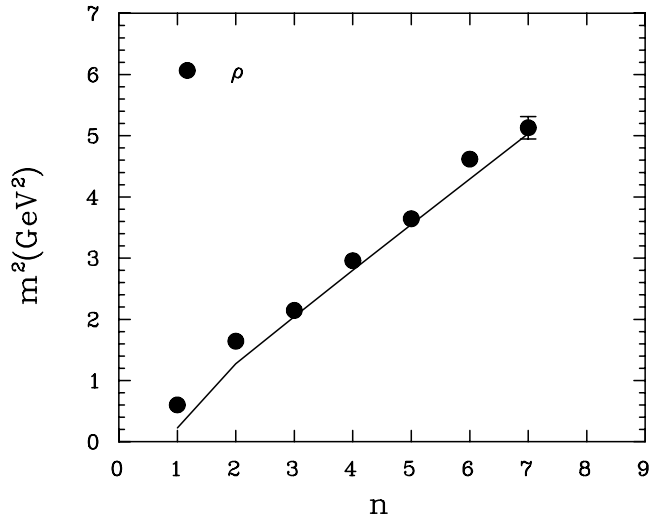}
\caption{\textsl{Comparison of the predicted vector mass eigenvalues using the tanh form (\ref{arcv}) of $v(z)$ (solid) with the QCD $\rho$ mass spectrum \cite{pdg}.}}
\label{VectorMasses}
\end{center}
\end{figure}

From the action (\ref{action1}) the equation of motion of the vector field can be written in the Schr\"{o}dinger form
\begin{equation} \label{schroedinger}
-\partial_z^{2}v_{n}+\left(\frac{1}{4}\omega'^{2} - \frac{1}{2}\omega''\right) v_{n}=m_{V_n}^{2}v_{n}
\end{equation}
where $\omega = \phi(z)+\log z$.  Using the dilaton form $\phi = \lambda z^{2}$, the eigenvalues of (\ref{schroedinger}) at large $n$ can be solved analytically with the boundary conditions $\lim_{z_0\rightarrow 0} v_n(z_0)=0$, $ \partial_z v_{n}(z\rightarrow\infty)=0$ 
and agree with those found in Ref. \cite{Karch:2006pv}.  However, since the dilaton specified in (\ref{phieqn}) is modified for $z\lesssim 1$ there is a change in the slope of the mass spectrum around $n=2$ which matches the behavior of the experimental data. The numerical vector mass spectrum is compared to the experimental data in Fig. \ref{VectorMasses}. While the prediction for the $\rho(775)$ mass is low, the rest of the vector meson masses are in reasonable agreement with experiment.  Most likely the agreement with the $\rho(775)$ could be improved upon by using a parameterization of $v(z)$ which rises more rapidly to its asymptotic value at large $z$.  Nevertheless, the purpose of this paper is to incorporate QCD-like chiral symmetry breaking in soft-wall AdS/QCD models, not just to fit data.

\subsection{Axial-vectors} 
\label{secAxial}

Unlike the vector field, the axial-vector couples to the scalar field VEV, producing a $z$-dependent mass term in its equation of motion.  
\begin{equation} \label{swaxialtrans}
-\partial_z^{2}a_{n}+\left(\frac{1}{4}\omega'^2 - \frac{1}{2}\omega'' +g_5^2 
\frac{L^2}{z^2} v^2(z) \right)a_n = m_{A_n}^2 a_n.
\end{equation}
The expression (\ref{swaxialtrans}) for the axial-vector field matches that of the vector field except for the additional term, $g_5^2 v^2(z) L^2/z^2$. Using the boundary conditions $\lim_{z_0\rightarrow 0} a_n(z_0)=0$, $ \partial_z a_{n}(z\rightarrow\infty)=0$, the axial-vector meson spectrum is obtained for the fixed values of $\lambda$, $m_q$, $\Sigma$, and $\kappa$. 

\begin{figure}
\begin{center}
\includegraphics[scale=0.40,angle=90]{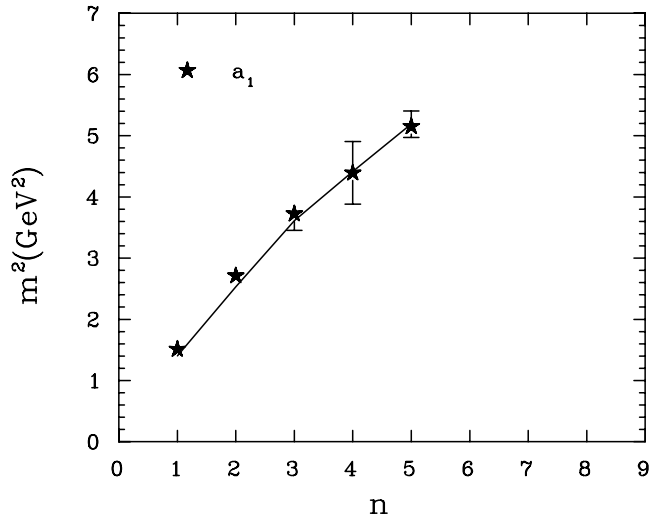}
\caption{\textsl{Comparison of the numerical results for the axial-vector mass eigenvalues using the tanh form (\ref{arcv}) of $v(z)$ (solid)
with the QCD $a_1$ mass spectrum \cite{pdg}.}}
\label{AxialMasses}
\end{center}
\end{figure}

The limiting behavior of $v(z)$ as $z\rightarrow\infty$ leads to a constant shift between the vector and axial-vector spectra at high mass values.  Comparing the equations of motion (\ref{schroedinger}) and (\ref{swaxialtrans})
for these fields one finds the asymptotic behavior
\begin{equation}
\Delta m^2 \equiv \left(m_{A_n}^2 - m_{V_n}^2\right)_{n \rightarrow \infty}
= g_5^2 \frac{L^2}{z^2} v^2(z \rightarrow \infty) = \frac{4g_5^2 \lambda}{\kappa}.
\label{deltam2}
\end{equation}
Together with the slope $\lambda$, this determines the numerical value to be $\kappa\sim 30$, although the best visual global fit to all the data suggests $\kappa = 15$, which is the value used here.  This is probably due to the small number of radial excitations to which we are fitting.  The results of our analysis are plotted in Fig. \ref{AxialMasses}. The $a_1(1260)$ resonance is predicted to within 5$\%$ and there is good agreement with the higher resonances of $a_1$.

Note that from (\ref{deltam2}), $\Delta m^2 >0$ implies that $\kappa >0$, which means that the potential in (\ref{action1}) is unbounded from below.  To address the stability of the gravity-dilaton background requires a complete fluctuation analysis generalizing the work in Ref. \cite{bl}. Even though this is beyond the scope of the present work it does suggest that higher-order terms will be needed for stability.

\section{Pion Coupling}
\label{secPion}

To confirm that our model is consistent with AdS/QCD model predictions, we calculate the pion decay constant using the formula \cite{Erlich:2005qh}
\begin{equation}
f_\pi^2=-\frac{1}{g_5^2} \lim_{\epsilon\rightarrow 0}
\frac{\partial_z A_0(0,z)}{z}\Bigg |_{z=\epsilon} \, .
\end{equation}
Here $A_0(q, z)$ is the axial-vector bulk-to-boundary propagator with boundary conditions $A_0(0,\epsilon)=1$ and $\partial_z A_0(0,z \rightarrow \infty)=0$. The above formula returns the measured value of the pion decay constant if $m_q = 9.75$ MeV, $\kappa = 15$, and $\lambda = 0.1831$ GeV$^2$.

\begin{figure}
\begin{center}
\includegraphics[scale=0.40,angle=90]{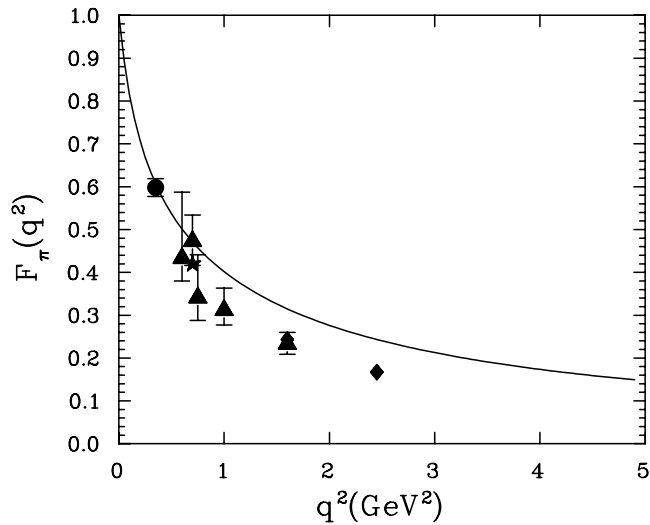}
\caption{\textsl{The line shows the predicted space-like behavior of the pion form factor $F_{\pi}(q^2)$ which is compared to the experimental data obtained from \cite{Kwee:2007nq}. The triangles are data from DESY, reanalyzed by \cite{Tadevosyan:2007yd}. The diamonds are data from Jefferson Lab \cite{Horn:2006tm}. The circles \cite{Ackermann:1977rp} as well as the star \cite{Brauel:1977ra} are also data obtained from DESY.}}
\label{Fpiplot}
\end{center}
\end{figure}

Once we calculate the $g_{\rho_{n}\pi\pi}$, the space-like pion form factor can easily be determined from a sum over vector meson poles. However the sum converges slowly and numerically it is much better to use the expression in terms of the vector and axial-vector bulk-to-boundary propagators as in Ref. \cite{Kwee:2007nq}
\begin{equation}
F_{\pi}(q^{2}) =\int{dz\, e^{-\phi(z)} \frac{V_0(q,z)}{f_{\pi}^{2}}\left(\frac{1}{g_5^2 z} (\partial_{z}\varphi(z))^{2} +\frac{v^2(z)}{z^{3}}(\pi(z)-\varphi(z))^2\right)},
\end{equation}
where $V_0(q,z)$ is the vector bulk-to-boundary propagator.
The results of our $F_{\pi}(q^{2})$ calculation are plotted in Fig. \ref{Fpiplot}, and shows a slight improvement in matching the experimental values compared to that obtained in Ref. \cite{Kwee:2007nq}.

\section{Conclusion} 
\label{secDiscuss}

We have shown how to incorporate chiral symmetry breaking into a soft-wall version of the AdS/QCD model with independent sources for spontaneous and explicit breaking. This is achieved by introducing a quartic term in the potential for the bulk scalar field dual to the quark bilinear operator ${\bar q} q$. This enables us to obtain reasonable agreement within 10$\%$ of the QCD meson mass spectra for scalar, vector and axial-vector fields, although the prediction for the lowest lying $\rho$ is not as good.

Further progress must recognize the limitations of this type of phenomenological model. Genuine stringy behavior is most likely required to fully describe the characteristics of QCD. On the theoretical side it would be interesting to further understand the soft-wall model from the top-down including finding a dynamical solution of the features exhibited in our model along the lines considered in Ref. \cite{Batell:2008zm}.  In addition the stability of the scalar potential will most likely require higher-order terms that can only be studied from the top-down. It is interesting that the simple 5D model contains QCD-like features and suggests that a further understanding of QCD can be obtained from the gauge/gravity correspondence.

\section*{Acknowledgments}
This work was supported by the Research Corporation for Science Advancement, the Australian Research Council, the US Department of Energy under grant 
DE-FG02-87ER40328, and by the School of Physics and Astronomy at the University of Minnesota.


\begin{thebibliography}{99}

\bibitem{Maldacena:1997re}
J. M. Maldacena, Adv. Theor. Math. Phys. {\bf 2}, 231 (1998)
[Int. J. Theor. Phys. {\bf 38}, 1113 (1999)].
  
\bibitem{Gubser:1998bc}
S. S. Gubser, I. R. Klebanov and A. M. Polyakov,
Phys. Lett. {\bf B428}, 105 (1998).
  
\bibitem{Witten:1998qj}
E. Witten, Adv. Theor. Math. Phys. {\bf 2}, 253 (1998).

\bibitem{Klebanov:1999tb}
I. R. Klebanov and E. Witten,
Nucl. Phys. {\bf B556}, 89 (1999).
  
\bibitem{Erlich:2005qh}
J. Erlich, E. Katz, D. T. Son and M. A. Stephanov,
Phys. Rev. Lett. {\bf 95}, 261602 (2005).
  
\bibitem{DaRold:2005zs}
L. Da Rold and A. Pomarol,
Nucl. Phys. {\bf B721}, 79 (2005).
  
\bibitem{Karch:2006pv}
A. Karch, E. Katz, D. T. Son and M. A. Stephanov,
Phys. Rev. D {\bf 74}, 015005 (2006).
  
\bibitem{Shifman:2007xn}
M. Shifman and A. Vainshtein,
Phys. Rev. D {\bf 77}, 034002 (2008).

\bibitem{us}
T. Gherghetta, J. I. Kapusta and T. M. Kelley, Phys. Rev. D {\bf 79}, 076003 (2009). 

\bibitem{Evans:2006ea}
N. Evans and A. Tedder,
Phys. Lett. {\bf B642}, 546 (2006).

\bibitem{Grigoryan:2007my}
H. R. Grigoryan and A. V. Radyushkin,
Phys. Rev. D {\bf 76}, 095007 (2007).
  
\bibitem{Kwee:2007nq}
H. J. Kwee and R. F. Lebed,
Phys. Rev. D {\bf 77}, 115007 (2008).
  
\bibitem{Cherman:2008eh}
A. Cherman, T. D. Cohen and E. S. Werbos,
Phys. Rev. C {\bf 79}, 045203 (2009).

\bibitem{Colangelo:2008us}
P. Colangelo, F. De Fazio, F. Giannuzzi, F. Jugeau and S. Nicotri,
Phys. Rev. D {\bf 78}, 055009 (2008).

\bibitem{Batell:2008zm}
B. Batell and T. Gherghetta,
Phys. Rev. D {\bf 78}, 026002 (2008).
  
\bibitem{Huang:2007fv}
M. Huang, Q. S. Yan and Y. Yang,
arXiv:0710.0988

\bibitem{Cohen:2005am}
T. D. Cohen and L. Y. Glozman,
Mod. Phys. Lett. A {\bf 21}, 1939 (2006).
  
\bibitem{Wagenbrunn:2006cs}
R. F. Wagenbrunn and L. Y. Glozman,
Phys. Lett. {\bf B643}, 98 (2006).
  
\bibitem{pdg}
C. Amsler, {\it et al.}, [Particle Data Group],
Phys. Lett. {\bf B667}, 1 (2008).

\bibitem{Bugg}
D. V. Bugg,
Phys. Rept. {\bf 397}, 257 (2004).

\bibitem{Bertin}
A. Bertin, {\it et al.} [OBELIX Collaboration],
Phys. Lett. {\bf B414}, 220 (1997).
  
\bibitem{bl} 
P. Breitenlohner and D. Z. Freedman,
Phys. Lett. {\bf B115}, 197 (1982).

\bibitem{LinSerot}
W. Lin and B. D. Serot, Phys. Lett. {\bf B233}, 23 (1989).

\bibitem{Tadevosyan:2007yd}
V. Tadevosyan, {\it et al.},  [Jefferson Lab F(pi) Collaboration],
Phys. Rev. C {\bf 75}, 055205 (2007).

\bibitem{Horn:2006tm}
T. Horn, {\it et al.},  [Jefferson Lab F(pi)-2 Collaboration],
Phys. Rev. Lett. {\bf 97}, 192001 (2006).

\bibitem{Ackermann:1977rp}
H. Ackermann, {\it et al.},
Nucl. Phys. {\bf B137}, 294 (1978).
  
\bibitem{Brauel:1977ra}
P. Brauel, {\it et al.},
Phys. Lett. {\bf B69}, 253 (1977).

  
\end{thebibliography}
\end{document}